\documentclass[10pt,preprint]{emulateapj}

\def\t0{\theta_{\circ}}

\def\be{\begin{equation}}
\def\en{\end{equation}}

\def\msun{M_{\sun}}

\def\lsun{L_{\sun}}
\def\msunyr{M_{\sun} \, yr^{-1}}

\def\h2{H$_2$}

\begin{document}

\title
{Short Gas Dissipation Timescales: Diskless Stars in Taurus and Chamaeleon I}

\author{Laura Ingleby\altaffilmark{1}, Nuria Calvet\altaffilmark{1}, Gregory Herczeg\altaffilmark{2}, Cesar Brice{\~n}o\altaffilmark{3}
}

\altaffiltext{1}{Department of Astronomy, University of Michigan, 830 Dennison Building, 500 Church Street, Ann Arbor, MI 48109, USA; lingleby@umich.edu, ncalvet@umich.edu}
\altaffiltext{2}{The Kavli Institute for Astronomy and Astrophysics, Peking University, Yi He Yuan Lu 5, Hai Dian Qu, Beijing 100871, P. R. China}
\altaffiltext{3}{Visiting Scientist, Department of Astronomy, University of Michigan, 830 Dennison Building, 500 Church Street, Ann Arbor, MI 48109, USA}

\begin{abstract}
We present an Advanced Camera for Surveys/ Solar Blind Channel far-ultraviolet (FUV) study of \h2 gas in 12 weak T Tauri stars in nearby star-forming regions. The sample consists of sources which have no evidence of inner disk dust.  Our new FUV spectra show that in addition to the dust, the gas is depleted from the inner disk.  This sample is combined with a larger FUV sample of accretors and non-accretors with ages between 1 and 100 Myr, showing that as early as 1--3 Myr, systems both with and without gas are found.  Possible mechanisms for depleting gas quickly include viscous evolution, planet formation and photoevaporation by stellar radiation fields.  Since these mechanisms alone cannot account for the lack of gas at 1--3 Myr, it is likely that the initial conditions (e.g. initial disk mass  or core angular momentum) contribute to the variety of disks observed at any age.  We estimate the angular momentum of a cloud needed for most of the mass to fall very close to the central object and compare this to models of the expected distribution of angular momenta.  Up to 20\% of cloud cores have low enough angular momenta to form disks with the mass close to the star, which would then accrete quickly; this percentage is similar to the fraction of diskless stars in the youngest star forming regions.  With our sample, we characterize the chromospheric contribution to the FUV luminosity and find that $L_{FUV}/L_{bol}$ saturates at $\sim10^{-4.1}$.

\end{abstract}


\section{ Introduction}
\label{intro}
An early stage of pre main-sequence evolution is characterized by the accretion of gas from the inner disk by magnetospheric
accretion.  The inner disk is truncated by the
stellar magnetic field and infalling gas is channeled onto the star by the
field lines creating a shock upon impact with the photosphere \citep{calvet98}.  As the source evolves the mass accretion rate decreases, until eventually, accretion ends when the inner disk gas is depleted (Ingleby et al. 2009; hereafter I09).  Accreting young stars with gas rich disks are called classical T Tauri stars (CTTS)
and non-accreting T Tauri stars with gas poor disks, weak T Tauri stars (WTTS).  Little is known about  how the gas is
ultimately depleted from the disk, marking the transition from CTTS to WTTS.   The amount of gas remaining in the disk may affect the eccentricity and migration of planetary orbits \citep{kominami02,matsuyama03} and therefore the gas dissipation timescale is key for models of planet formation and evolution.  

Circumstellar disk evolution is primarily
traced by dust in the disk, emitting at infrared (IR) wavelengths \citep{hernandez08}.  Attempts at modeling the IR excess have resulted in a possible sequence of disk evolutionary stages.  Beginning
with a full disk of small dust particles distributed
throughout the vertical layers, the dust coagulates and
settles towards the midplane
\citep{weidenschilling97,dullemond04}.  There, a planet may form, and if massive enough, open a gap in the disk
\citep{zhu11}, observed as a deficit in IR fluxes.  A disk gap may also form when the mass loss rate  due to photoevaporation of disk material by high radiation from the central star exceeds the mass accretion rate  \citep{clarke01}.  After the gap is formed the inner disk is 
drained onto the
star by viscous evolution and/or is removed from the system through continued
photoevaporation, creating an inner
disk hole.

Circumstellar gas evolution has proved harder to probe than the dust, due to the difficulty of detecting spectroscopic gas signatures.  Recently, observations of molecular gas in the disks of low mass T Tauri stars have increased, including IR detections with the \emph{Spitzer} Infrared Spectrograph (IRS) (Carr \& Najita 2011, and references therein),  \emph{Herschel} (e.g. Mathews et al. 2010) and \emph{Hubble Space Telescope} (HST) FUV observations of \h2 \citep{france11,ingleby11b,herczeg02,calvet04,yang12}.  In a study of 43 young stars, I09 used low resolution FUV spectra from the Advanced Camera for Surveys Solar Blind Channel (ACS/SBC) PR130L prism to study gas in disks surrounding accreting and non-accreting sources between 1 Myr and 1 Gyr.  I09 located a continuum feature in the low resolution spectra near 1600 {\AA} produced when \h2 is excited by high energy electrons \citep{bergin04} and found it present in accreting sources but absent in
non-accreting sources.  Since the X-ray emission
necessary for producing the high energy electrons is strong during pre-main sequence evolution \citep{ingleby11a}, the lack
of \h2 emission in the non-accretors indicates that gas is cleared from the inner disk by the time accretion ends.  However, I09 only observed non-accreting sources at ages $>$10 Myr and observations of young WTTS were necessary to confirm that the gas is completely drained, even when accretion stops early on.

We present FUV observations of 1--3 Myr WTTS in the Taurus and Chamaeleon I star forming regions and look for evidence of circumstellar \h2.  In Section \ref{results} we present results on the presence or lack of \h2 and in Section \ref{disc} we discuss mechanisms which may be responsible for disk evolution in these sources, as well as the evolution of high energy radiation from the star.

\section{Sample, Observations and Data Reduction}
\label{obs}

\subsection{FUV Observations}
We obtained observations centered on 11 young stars using ACS/SBC on \emph{HST} in GO program 12211 (PI: Calvet).  Two of the fields contained wide binaries where the spectrum of each component was extracted separately; however, one of the single stars was not detected in the FUV, therefore the total sample consists of 12 sources. Each ACS observation consists of a brief image in the F165LP filter and a longer image obtained with the PR130L prism. Offsets between the target location in the filter and prism image, including the wavelength solution, were obtained from \citet{larsen06}. The target spectrum was then extracted from a 41-pixel (1\farcs3) wide extraction window. Background count rates were calculated from offset windows and subtracted from the extracted spectrum. The absolute wavelength solution was then determined by fitting the bright C IV $\lambda$1549 {\AA} doublet. Fluxes were calibrated from the sensitivity function obtained from white dwarf standard stars by \citet{bohlin07}. The spectra cover 1230--1800 {\AA} with a 2-pixel resolution of  $\sim$300 at 1230 {\AA} and $\sim$80 at 1600 {\AA}.

\begin{figure}
\plotone{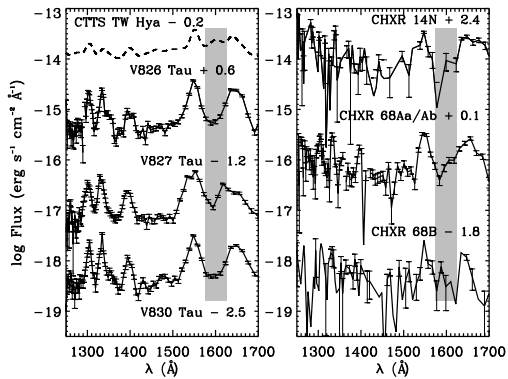}
\caption{ACS/SBC observations.  The spectra are offset by the value listed.  A Space Telescope Imaging Spectrograph spectrum of the CTTS TW Hya is shown for comparison (Herczeg et al. 2002; dashed line), convolved to the PR130L resolution.  The grey shaded area represents the spectral region where we would expect to see \h2 emission, if gas were present in the disk (I09); in TW Hya \h2 emission is observed.  For the Chamaeleon I sources, it is unclear if emission near 1600 {\AA} is produced by \h2 or noise in the spectrum as the error bars are large in this region; the signal to noise in the continuum is $<$3, as opposed to $>$5 for the Taurus sample.}
\label{acs}
\end{figure}

\subsection{Notes on Sample and Individual Sources}
The FUV detections include 7 low mass T Tauri stars in Taurus and 5 in Chamaeleon I (Figure \ref{acs} and Table 1).  WTTS were chosen based on a lack of evidence for remaining inner disk dust as traced by $K-L < 0.4$ for Taurus sources (Kenyon \& Hartmann 1995), or by a Spitzer IRS slope $< -2.2$ between 2 and 24$\mu$m or 3.6 and 24$\mu$m for sources in Chamaeleon I (Luhman et al. 2008), and no ongoing accretion, defined as H$\alpha$ equivalent width (EW) $<$ 10 {\AA} for the spectral types of this sample (Kenyon \& Hartmann 1995; Luhman 2004).  Very low accretion may not register on the H$\alpha$ line profile in low resolution spectra \citep{ingleby11b}; therefore, small quantities of gas could remain in the inner disk.

We detect both components of LkCa 3, a binary with separation of 69 AU  \citep{kraus11}.  We use a spectral type of M1 and assume equal contribution to the unresolved luminosity from each \citep{kenyon95}.  For V827 Tau, the lines are too broad, indicating a binary aligned along the dispersion direction.  \citet{kraus11} found that V827 Tau is a close binary with mass ratio of $\sim$0.6 and separation of $\sim$90 mas (13.5 AU).  With the close separation we do not separate the two components.  CHXR 68 is a triple system with A and B components separated by 4\farcs4 and the A component itself a binary with a separation of 0\farcs1 and mass ratio of 0.6 \citep{nguyen12}.  We extract A and B individually and identify the two components of the A binary with a separation of 60 mas, but again due to the close separation, the prism spectrum of CHXR 68A is a combination of the two sources.  J1-4827 was not detected in the FUV.

\begin{figure}[htp]
\plotone{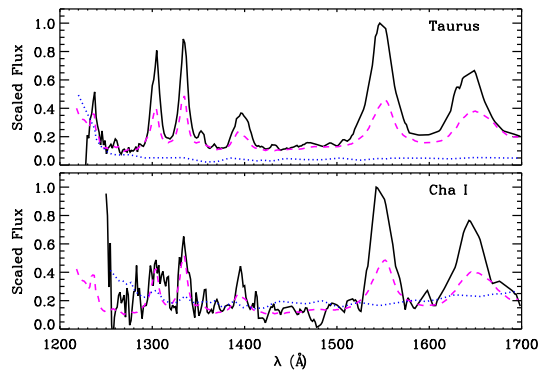}
\caption{Taurus and Chamaeleon I ACS spectra.  We compare the median of the Taurus (top panel) and Chamaeleon I (bottom panel) WTTS observations (black line) to the median of the WTTS/DD sample observed in I09 (dashed purple line).  We also show the level of the background (dotted blue line).  For Chamaeleon I, the background emission is comparable to the source flux at 1600 {\AA}.  }
\label{median}
\end{figure}

\section{Results}
\label{results}

We used a continuum feature at 1600 {\AA}, which is formed primarily by the photodissociation of 
\h2, as a gauge of the strength of \h2 emission  \citep{bergin04}.  We first de-reddened the spectra using published $A_V$ or $A_J$ magnitudes \citep{kenyon95,luhman04} and the \citet{whittet04} extinction law.  The FUV continuum{\footnote{Due to the low resolution of the ACS/SBC prism, the observed continuum includes unresolved line emission blended with the intrinsic continuum from the source.}}  has contributions from several sources in CTTS, including the chromosphere and accretion shock, in addition to the continuum produced by \h2 dissociation \citep{ingleby11a,ingleby11b}.  To approximate the FUV continuum at the PR130L resolution, we used the median FUV spectrum of the WTTS and debris disks (DD) in I09.  I09 showed that no \h2 emission was present in the spectra of these sources and there is no ongoing accretion, leaving only the chromospheric contribution.  Figure \ref{median} compares the median of the I09 WTTS/DD sample to the median of the Taurus and Chamaeleon I WTTS samples.   After subtracting the I09 WTTS/DD spectrum, scaled to the target spectra between 1400 and 1500 {\AA}, the flux between 1575 and 1625 {\AA} was integrated and represents the flux of the \h2 feature. 

\begin{figure}
\plotone{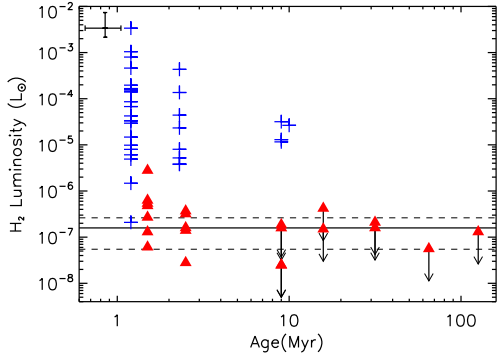}
\caption{\h2 versus age.  The luminosity of the \h2 feature for accreting sources is shown as blue crosses and for the non-accretors, red triangles.  New WTTS in Taurus and Chamaeleon I are offset from the CTTS in each region by 0.2 Myr for clarity.  The solid line represents the median \h2 luminosity for the I09 non-accreting sample and the dashed lines represent 3$\sigma$ from that median.  Typical errors are shown in the upper left corner.}
\label{h2}
\end{figure}

Figure \ref{h2} shows the luminosities of the \h2 feature in our sample combined with results from I09.  We also include new observations of accretors in Chamaeleon I \citep{ingleby11a}.  Many of the 1--3 Myr WTTS have \h2 feature luminosities within 3$\sigma$ of the luminosities of the I09 older population which were shown to be cleared of gas in the inner disk.   The \h2 feature luminosities of the WTTS sample are lower than all but two of the Taurus accretors.  For V827 Tau, the luminosity is high ($>10^{-6}\;\lsun$) because of contributions from wide C IV and He II emission lines produced by the binary observed along the dispersion axis.  The CTTS HN Tau B has an \h2 luminosity comparable to our WTTS sample; it is a low mass companions with spectral type M4 and unknown accretion properties \citep{kraus11}.  Given the large error on the low resolution spectra and the similarity in \h2 feature luminosities with the I09 sample, these results are consistent with the 1--3 Myr sample of WTTS being cleared of inner disk gas.

\section{Discussion}
\label{disc}

\subsection{Disk Depletion on Short Timescales}
As shown in Figure \ref{h2}, circumstellar gas is depleted in short timescales for some sources, while others retain a gas disk, even at 10 Myr.  The diskless sources at 1--3 Myr are interesting because they have lost their disks in timescales much shorter than the typical disk lifetime, 5--10 Myr \citep{hernandez08,fedele10}.  The reason no disks remain for these sources may be due to the initial conditions, e.g. the initial disk mass or angular momentum of the parent cloud \citep{shu87}, or may stem from the physical processes which are responsible for dispersing disk material.  Close binaries ($<$ 40 AU) may be incapable of retaining a disk; however only three of our sources are close binaries, V826 Tau, V827 Tau and CHXR 68A \citep{kraus12,nguyen12}.  Other possible mechanisms for disk dispersal include viscous evolution and accretion of gas, planet formation and photoevaporation.

In addition to gas, the dust is gone, based on photospheric IR fluxes observed in IRS spectra indicating a lack of inner disk dust \citep{luhman08,wahhaj10}.  \citet{andrews05} observed five of the Taurus sources in the sub-mm, tracing the outer disk, and constrained the disk mass to $<$0.0004 $\msun$ ($<2\;M_{Jup}$).   They also compared sub-mm and IR data for their sample and found that most sources lacked excesses in both, indicating that inner and outer disks are dispersed within 10${^5}$ years of each other.  Therefore, both the gas and dust in both the inner and outer disks of our sample were likely depleted quickly.   While planets alter the disk structure by forming a gap at their orbital radius, it is unlikely that they will completely disperse both the inner and outer disks \citep{zhu11}.  

Blue-shifted forbidden line emission at velocities around 10 km s$^{-1}$ is interpreted as evidence that gas is undergoing photoevaporation \citep{pascucci11}.  Theories of photoevaporation differ regarding the type of high energy emission responsible for the mass loss and the rates at which the material is dispersed, yet all models eventually produce a gap in the disk.  Initial models of photoevaporation by EUV emission predicted low mass loss rates of $\sim10^{-10}\;\msunyr$ \citep{font04,alexander06}.  \citet{alexander06} showed that once the initial gap is formed, both the inner and outer disks disappear quickly, in $\sim0.1-0.2$ Myr, consistent with the observations of \citet{andrews05}.  After the gap is formed, the inner and outer disk are influenced by continued accretion and photoevaporation, respectively, at rates of $10^{-10}\;\msunyr$.  At these low rates, the disk mass at the time of gap opening must be less than $2\times10^{-5}\;\msun$ in order for the disk to be depleted in $10^{5}$ years.  Given the early age of the sources in our sample, there is little time for the disk to be depleted down to the low masses needed for EUV photoevaporation to disperse the disk.

A larger disk mass is possible if the mass loss rate is higher than that predicted by EUV photoevaporation, and recent models of photoevaporation which include X-ray and FUV emission achieve mass loss rates up to $10^{-8}\;\msunyr$ \citep{gorti09,owen11}.   In \citet{owen11}, the mass loss rate scales with the X-ray luminosity; for the X-ray luminosities of our WTTS sample (see $\S$4.2), the predicted mass loss rate would be $\sim10^{-8}\;\msunyr$.  Therefore both the inner and outer disk would contain $<10^{-3}\;\msunyr$ to be completely removed in $<10^5$ years,  close to the range of predicted disk masses.  However, objects with very low accretion rates yet with full disks cannot be undergoing photoevaporation at such high rates, whereas the low mass loss rates predicted by the EUV photoevaporation models are more consistent with observations \citep{ingleby11b}. 

Instead, WTTS may form from cloud cores with low initial angular momentum, $J$.  Assuming stars both with and without disks at 1-3 Myr formed from cloud cores with the same mass, the cloud angular momentum determines the mass distribution in the disk.  The radius at which the cloud material accretes onto the disk is given by $r_c$, which depends on the core gas temperature ($T$), angular velocity ($\Omega$), and mass of the central star ($M$).  Assuming $T=10$ K and $M=1\;\msun$, 
\begin{equation}
r_c\sim9\;\rm{AU}\times\Omega^2_{-15}
\end{equation}
where $\Omega_{-15}$ is in units of $10^{-15}$ rad s$^{-1}$ \citep{hartmann09}.  Assuming uniform rotation, the specific angular momentum is 
\begin{equation}
J/M\sim 9.5\times10^{19}\; \rm{cm}\; \rm{ s}^{-2\;} R_{0.1}^2\;\Omega_{-15}
\end{equation}
where $R_{0.1}$ is the initial radius of the cloud in units of 0.1 pc.  If matter falls from a cloud with $R_{0.1}$=0.1 pc at $r_c\le$ 1 AU (which would accrete onto the star in only 30,000--40,000 years), then J/M=19.5 cm$^2$ s$^{-1}$.  This is a rough approximation of the angular momentum, as the actual collapse of cloud cores is significantly more complicated than assumed here and physical properties of the cloud (for example, the magnetic field strength) may be important \citep{allen03}.  Up to 20\% of cloud cores have low enough angular momenta to form a disk which would accrete quickly, according to the models of \citet{dib10}.  The observed disk fractions in the youngest clusters are 90--95\% \citep{hernandez08}.  Given the distribution of angular momenta predicted by \citet{dib10}, it is possible that the 5--10\% of diskless stars formed from slowly rotating clouds.

\subsection{Evolution of Chromospheric Emission}

Accurate estimates of the high energy emission in young stars are important because one or a combination of these fields may drive photoevaporation.  The active chromospheres in young stars produce a UV excess with respect to the photosphere \citep{houdebine96}; however, the total excess is difficult to estimate due to the small number of UV observations of WTTS.  Here, we analyze the FUV emission of 23 non-accreting young stars (combining our sample with the WTTS/DD sample in I09) to characterize the evolution of chromospheric emission.  WTTS have no emission contributions from accretion or \h2, so the FUV continuum and hot lines are intrinsic to the young star and provide a gauge of the chromospheric emission. 

\begin{figure}
\plotone{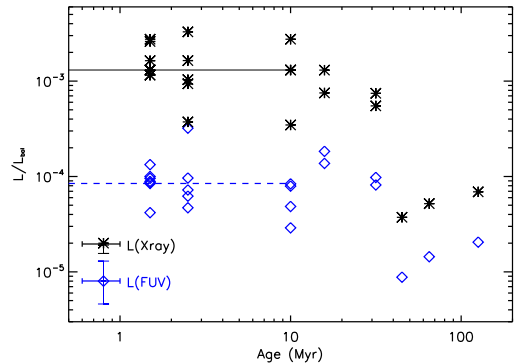}
\caption{X-ray (black asterisks) and FUV (blue diamonds) luminosities normalized by the bolometric luminosity versus age.   The median of the X-ray and FUV luminosities for  sources at $\le$10 Myr are indicated with black solid and dashed blue lines, respectively.  Typical errors are shown in the lower left corner.}
\label{xrayfuv}
\end{figure}

We also include published X-ray observations to probe the evolution of coronal emission \citep{voges99,feigelson93,gudel07}.  ROSAT and XMM-Newton X-ray count rates were converted to X-ray luminosities in the 0.2-10 keV range using the HEASARC tool WebPimms for comparison to the X-ray luminosities of the I09 WTTS/DD sample.  In Figure \ref{xrayfuv}, we show the evolution of high energy stellar radiation fields between 1 and 100 Myr {\footnote{In \citet{ingleby11a}, one source appeared to diverge from the fit to the X-ray decline with age.  This source, HD 53143, has an uncertain age, between 45 Myr to 6 Gyr \citep{nakajima10,holmberg09}.  We find that the young age is more consistent with the observed decline in X-ray emission and therefore assume an age of 45 Myr.}}. Large X-ray samples of WTTS indicate that the emission saturates around 10 Myr \citep{preibisch05}.   The X-ray saturation level of our sample, log $L_X/L_{bol} = -2.9\pm0.2$, is within the errors of that found previously,  log $L_X/L_{bol} = -3.3\pm0.4$.

The evolution of FUV emission is not as well studied as the X-ray evolution.  \citet{ribas05} observed that high energy emission, including FUV, decreases for solar type stars $>$ 100 Myr.  \citet{findeisen11} determined the chromospheric indicator $R^{'}_{FUV}$ in a sample with GALEX data and observed a decrease in chromospheric activity between $10^{7.5}$ and $10^9$ years.  Evidence suggests that the same physical mechanism is heating both the chromosphere and the corona; \citet{maggio87} and more recently \citet{mamajek08} found that the coronal X-ray luminosity was correlated with indicators of the chromospheric emission  (e.g. the Ca II H and K lines) and \citet{ingleby11a} showed that X-ray and FUV luminosities are correlated for $>10$ Myr non-accretors.  Given this evidence that the heating of the chromosphere and corona are linked and should therefore saturate on similar timescales, we estimate the FUV saturation for sources $\le$ 10 Myr (shown in Figure \ref{xrayfuv}.)  Total FUV luminosities were integrated between 1230 and 1800 {\AA} (Table 1).  We find that the FUV luminosity saturates at a level of log $L_{FUV}/L_{bol}=-4.1\pm0.1$. We also see evidence for the beginning of the expected decline in FUV emission for the oldest sources in our sample.

\subsection{Summary}
We presented results from an FUV study of the gas (or lack of gas) in the disks around 1-3 Myr WTTS.  Our conclusions are summarized here.
\begin{itemize}
\item
A new FUV sample of young (1--3 Myr) WTTS in the Taurus and Chamaeleon I star forming regions confirms previous results that \h2 gas in the inner circumstellar disk is completely dissipated by the time accretion onto the star ends.  This is true even for the youngest non-accretors.
\item
Photoevaporation alone cannot explain diskless stars at 1--3 Myr, unless the initial disk mass is very small.  However, up to 20\% of T Tauri stars may form from clouds with low angular momenta, producing disks with small radii which accrete quickly onto the star.
\item
If the heating of the corona and chromosphere are related, as observational evidence indicates, we expect the X-ray and FUV emission to saturate at the same age.  The saturation of the FUV emission for our sample occurs at log $L_{FUV}/L_{bol}=-4.1\pm0.1$.  
\end{itemize}

\section{Acknowledgments}
This work was supported by NASA grants for Guest Observer program 12211 to the University of Michigan.


\begin{deluxetable}{llcccc}
\tablewidth{0pt}
\tablecaption{Source Properties}
\label{tabprop}
\tablehead{
\colhead{Object} &\colhead{SpT}& \colhead{Luminosity}&\colhead{$A_V$}&\colhead{L$_{FUV}$}& \colhead{Region}\\
\colhead{} &\colhead{}& \colhead{(${\lsun}$)}&\colhead{}&\colhead{($\times10^{-5}\;\lsun$)}& \colhead{} }  
\startdata
J1-4872&K7&0.5&0.0&--&Taurus\\
LkCa 3E&M1&0.9&0.4&8.0&Taurus\\
LkCa 3W&M1&0.9&0.4&8.6&Taurus\\
LkCa14&M0&0.4&0.0&3.4&Taurus\\
V826 Tau&K7&0.8&0.3&6.8&Taurus\\
V827 Tau A/B&K7&0.8&0.3&8.0&Taurus\\
V830 Tau&K7&0.6&0.3&8.0&Taurus\\
V1075 Tau&K7&0.5&0.0&2.1&Taurus\\
CHXR 14N&K8&0.4&0.5&2.5&Cha I\\
CHXR 55&K5&0.7&1.1&22.5&Cha I\\
CHXR 68 Aa/Ab&K8&0.7&0.4&3.3&Cha I\\
CHXR 68 B&M2&0.2&0.5&1,9&Cha I\\
UV Cha&M2&0.4&0.4&2.9&Cha I
\enddata
\tablecomments{Spectral type, luminosity and $A_V$ were taken from \cite{kenyon95} and \citet{luhman04} for Taurus and Chamaeleon I, respectively.}
\end{deluxetable}

\end{document}